# Software Validation using Power Profiles

**Raimondas Lencevicius      Edu Metz      Alexander Ran**

*Nokia Research Center, 5 Wayside Road, Burlington, MA 01803, USA*
*Raimondas.Lencevicius@nokia.com    Edu.Metz@nokia.com    Alexander.Ran@nokia.com*

**Abstract**. *The validation of modern software systems incorporates both functional and quality requirements. This paper proposes a validation approach for software quality requirement—its power consumption. This approach validates whether the software produces the desired results with a minimum expenditure of energy. We present energy requirements and an approach for their validation using a power consumption model, test-case specification, software traces, and power measurements. Three different approaches for power data gathering are described. The power consumption of mobile phone applications is obtained and matched against the power consumption model.*

## 1   INTRODUCTION

While historically software validation focused on the functional requirements, recent approaches also encompass the validation of quality requirements; for example, system reliability, performance or usability. However, application development for mobile platforms opens an additional area of quality—power consumption. In PDAs or mobile phones, power consumption varies depending on the hardware resources used, making it possible to specify and validate correct or incorrect executions. Consider the following example: Assume a simple device model of CPU, display and network that use some power when active and use zero power otherwise. An application downloads a video stream from the network and displays it on the mobile device's display. In the test scenario the viewing of the video is paused at a certain point. If the specification does not allow video prefetching for caching on the mobile device, the user expects that the network card activity would stop when the video is paused. How would a test engineer check this expectation? Simply running a test suite or even tracing the software execution does not detect the network activity, since a test suite usually reports only the results of the software execution and not all system activity. However, the extraneous network activity can be detected by power measurements and power model application (Figure 1). Power requirement violations such as described above are probable in embedded SW development, where developers have to deal with numerous different devices, each with its own protocol, requirements and software. Tools to find the power inconsistencies and to validate software from the energy point of view are needed.

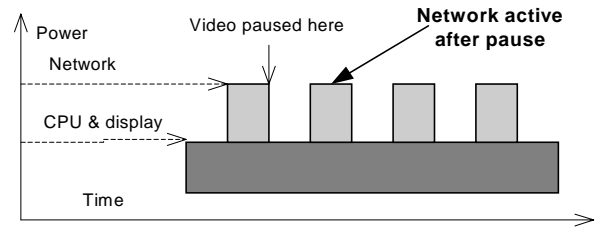

**Figure 1. Hypothetical streaming video power profile**

This initial concept paper proposes an energy and power validation approach. First, we argue that such validation is useful, especially in mobile device domain. Section 3 explains energy requirements, the energy and power consumption models, test-case specifications and the energy validation approach. Section 4 presents different approaches to power data collection. Section 5 describes our power measurement framework. The paper discusses energy and power validation of mobile phone applications in section 6. We finish with related work and conclusions.

Since power is energy expended over a period of time ($P = \Delta E/\Delta t$), we use the term "power" when we refer to instantaneous measurements or profiles composed from such measurements. We use "energy" to refer to the total energy spent over a period of time. As the formula above shows, these two quantities are easily calculated from each other.

## 2   MOTIVATION

Validating modern systems is done against functional and quality requirements. Test suites are designed to validate the implementation of a system against the requirements. Test suites provide an environment for the controlled execution of the system in which the validation can be performed. The test suite's results are usually compared against the final output of the system and perhaps against some events that occur during the program's execution. Most of these events are user interface events though in some areas other events such as network messages are monitored [8]. However, executing applications interact not only with display, but also with various other hardware devices: processor, memory, network interface, and possibly hard drive. All these devices consume energy during operation. This gives an opportunity to monitor and record power levels during the test suite execution. The recorded power levels can then be used to validate energy requirements.

Energy quality requirements specify that the software should execute with minimal power levels and energy consumption. Rapid growth of intelligent handheld devices requires the development of numerous software applications. Developing software applications for mobile devices is different than developing applications for desktop computers. Mobile devices interact with a much richer set of hardware. Mobile devices may use Bluetooth interface [2], display, cellular interface, flash memory, DSP chips, infrared communications and so on. In addition, mobile devices have limited processing power and memory as well as energy consumption constraints. Energy constraints are an important issue in software design for mobile devices. It is essential to design software to minimize energy consumption, preventing hardware accesses from depleting the battery of the mobile device faster than necessary. For this reason, extraneous hardware accesses that could be ignored in a desktop environment have to be checked much more closely in a mobile environment. We propose an approach to validate software from the energy point of view. The next section discusses the information needed for such validation.

## 3   ENERGY CONSUMPTION VALIDATION

The energy consumption validation has three parts: energy consumption requirements, device power model, and system observation.

Energy consumption requirements depend on a concrete software and hardware system. This paper outlines general intuitive requirements and mentions a few of the possible extensions. Let us model the device's software as a collection of run-time functions that are executed, possibly in parallel. Hardware devices are associated with software functions. The efficient energy consumption requirement then is that only hardware devices associated with active software functions should be active. This requirement relates to the example given in the introduction—when the video playback is paused, the active software function is not associated with the network card, so the network card should be inactive. Some devices do not follow the model above and therefore do not necessarily satisfy the above requirement. For example, there may be hardware devices controlled both by software and by hardware, e.g. speaker volume control that has a physical volume dial and a virtual scroll bar. Such device could be active even when there is no active software associated with it. In such cases the efficient energy consumption requirements are more complicated.

To validate the energy requirements, the device power model, power measurements and software traces are needed. We formalize and extend the power model proposed by Cignetti and others [4] by using state transition diagrams [3] and extended message sequence charts to model the power consumption.

We model the device under validation as a collection of hierarchical state machines representing power consuming components. The model can be build top-down or bottom-up. Top-down modeling starts from the whole device and then divides it into power consuming subsystems and components. In a bottom-up description, one starts from a set of device's atomic hardware components, for example, an LCD screen, a processor, a network interface and flash memory. Each such component is described by the state machine that contains a set of states with specified power functions. We call a hardware component "atomic" if it is not divided into smaller parts in a device model. Certain components together form device subsystems, for example, an audio subsystem, a display and so on. Subsystems are modeled as higher-level state machines containing nested state machines. Sometimes components may work in parallel, which is modeled as concurrent nested substates. Finally, the device itself is the collection of all state machines of its subsystems and components.

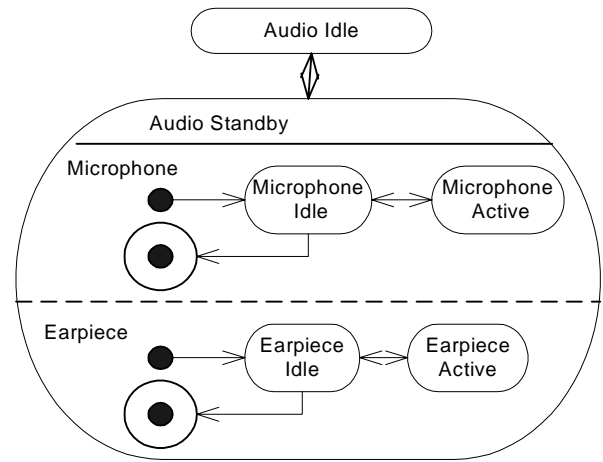

**Figure 2. Audio subsystem state transition model**

Figure 2 shows the state transition diagram of a simplified audio subsystem of a mobile device. It shows that the audio subsystem can be in an Audio Idle state or in an Audio Standby state that nests parallel substates describing microphone and earpiece activity. Both microphone and earpiece can become independently active or idle when the audio subsystem is in standby.

Each state and transition in the model—whether it is atomic or composite—is characterized by a power level function. In the simplest case, such function is a constant for a state and zero for a transition. For example, the display backlight consumes a constant amount of power when it is on, so the $P_{BacklightOn} = PConstant_{BacklightOn}$.

However, the power consumption in a state may depend on parameters. For example, the power consumption in a processor depends on executed instructions, cache misses, memory accesses, etc. The power consumption in a radio antenna depends on the transmission frequency, signal strength, transmission protocol and other parameters. It is possible to use only constant power functions and to model any parameter dependencies by introducing additional states and transitions. However, such approach is very cumbersome. For example, while a very detailed state transition diagram could potentially model the processor power consumption, such diagram would have enormous number of states and would be difficult to construct and understand. Non-constant power functions allow us to use higher-level system abstractions in state transition diagrams. In the processor example, it is simpler to use just one state for an active processor and a power function for this state that takes into account average instruction mix and cache miss ratio.

The power function of a composite state combines the power functions of the state's components. For example, the Audio Standby state's (Figure 2) power function combines the constant standby power of the audio subsystem $PConstant_{AudioStandby}$ and the power functions of the microphone $P_{microphone}$ and the earpiece $P_{earpiece}$:

$$P_{AudioStandby} = PConstant_{AudioStandby} \circ P_{microphone} \circ P_{earpiece}$$

In simple cases, the composite function is a sum of the component functions.

In the above-described model—the hierarchical state transition diagram—the transitions between the power states are triggered by global system events. Global system events apply to all concurrent parts of the state transition diagram, so when an event occurs, all such parts perform the state transitions triggered by this event. The power function of the system in each state determines the overall power consumption of the system.

The model above coupled with the observation of the global system events that trigger state transitions and with power-level measurements at the time of each event is sufficient for the energy requirement validation. Events are observed as software traces. Power measurements and software traces are described in the next section. During the validation the events from a trace are applied to the model in time order obtaining the state of the model and the corresponding modeled power level at each moment of time. This power level then is compared to the measured power to check the consistency between software traces, the power model and power measurements. If the modeled and the measured power levels differ, the consistency is violated.

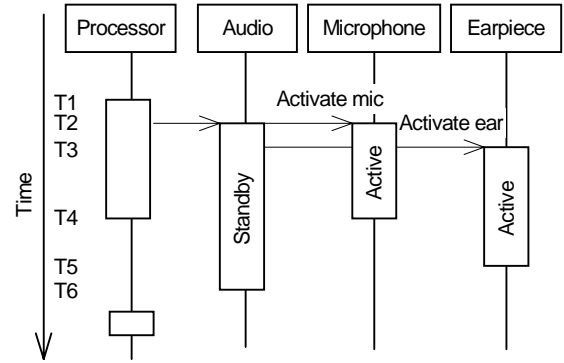

**Figure 3. Sequence chart specification for scenario involving audio system**

It is possible to obtain additional validation constraints by adding to our approach test-case specifications involving time. Such specifications can be modeled using extended message-sequence charts. Such message sequence charts not only determine the message and event sequence in a test case, but also specify the time constraints between events and messages similarly to timed transition systems [7] and modecharts [9]. Figure 3 shows an example of a test case involving a processor and audio subsystem of a device. The scenario starts at time T1 when the processor becomes active. At time T2 it sends a message putting the audio subsystem into a standby mode. At the same time the audio subsystem activates the microphone and at time T3 it activates the earpiece. At time T4 the processor and the microphone independently become idle, while the earpiece is active until time T5. The time constraints in specifications would not be expressed in absolute times, but in relative *delay-deadline* intervals [7]. For example, the interval (T1, T4) would be specified as the smallest—*delay*—and the largest—*deadline*—possible interval between the processor activation and shutdown. Assuming that the power composition of the subsystems and components is expressed as a sum, the energy specified in our example scenario is:

$$E_{Scenario} = \int_{T1}^{T4} P_{ProcessorActive}(t)dt + \int_{T2}^{T6} P_{AudioStandby}(t)dt + \int_{T2}^{T4} P_{MicrophoneActive}(t)dt + \int_{T3}^{T5} P_{EarpieceActive}(t)dt$$

Here $P_{ProcessorActive}(t)$, $P_{AudioStandby}(t)$, $P_{MicrophoneActive}(t)$, and $P_{EarpieceActive}(t)$ are the power level functions of subsystems in respective states. These functions may depend on additional parameters, such as sound volume and a melody. If the subsystems consume non-zero energy in idle states or in transitions, the integrals of the power functions of the idle states and transitions should also be added to the $E_{Scenario}$ function.

The test-case specifications allow to validate the completeness and correctness of event traces. Time intervals in the test-case specifications also specify additional constraints on the power model that can be checked through the power measurements. The power measurements can indicate that a certain subsystem was active shorter or longer than the time intervals of the test case allow. For systems with test-case specifications, the calculated energy can be compared to the actual energy consumption.

## 4 POWER DATA COLLECTION METHODS

The previous section described the formal approach for the energy validation. Such validation needs power measurements and global event tracing. This section describes three data collection approaches.

The first and simplest approach to gather data is to measure the power consumed by a device at regular time intervals. This approach produces a power level graph with no indications of what software was running and what hardware was accessed during the execution. Such graph may be useful in the software validation if tracing of interesting software cannot be achieved; for example, when the source code and symbol data are unavailable. Using the power model the graph can be deciphered by matching the power levels to the values given in the model. A test engineer may be able to identify graph segments corresponding to network activity, processor sleep states and writes to the flash memory. Then the annotated graph could be approximately matched to the power specification to determine any serious discrepancies. In addition this view can be used for a quick high-level validation. An experienced test engineer may see that the pattern of power levels is different than expected. Such view is similar to execution murals [10].

The more informative second approach still reports the power levels independently of the software execution on the system. However, the report also specifies the processes active at measurement points. Such approach is used at CMU in PowerScope setup [6]. This approach does not require explicit program instrumentation and yet maps software processes to their energy consumption. The accuracy of such mapping depends on the frequency of the energy readings, but usually achieves a per-procedure level. The drawback of PowerScope implementation is that it needs kernel modification of the underlying operating system and that it requires access to the symbol tables of the executables that were running during profiling.

Finally, we propose the third approach that combines tracing and the power measurements. Even though the traces by themselves yield a lot of information, the power measurements add an additional dimension to the trace information. First, the tracing may not cover all hardware access events. In such case the power measurements allow to detect holes in tracing instrumentation and then improve the tracing. Second, the power model may be incomplete, leading to insufficient tracing. In such case the measurements allow to improve the model and add additional traces connected to it. Third, the power measurements are used to calibrate the power functions and numerical values in the model. Examples of the power level measurements and modeling using the third approach in mobile device software validation are given in section 6.

Given that the last approach is the most powerful, are there any reasons to use one of the first two data gathering approaches? The first two approaches do not require the manual instrumentation of the source code. They provide a high-level view of the system's power levels and energy consumption, allowing coarse validation. They also help test engineers to identify suspicious areas and then concentrate on them with tracing. Ultimately, these three approaches are complementary. In our work, we have used both the first and the third data gathering approaches.

## 5 IMPLEMENTATION

We have implemented the data gathering for the energy validation in mobile phones using the first and the third approaches described in section 4. We measure the current drawn by a device with a digital multimeter. The multimeter is connected to the battery line of the mobile phone (Figure 4). The high-speed GPIB (IEEE-488) interface card plugs into a PCI slot of the PC and connects to multimeter via a GPIB cable.

Agilent BenchLink Meter software [1] acquires the measurement data with timestamps and exports it in "comma-separated-value" (CSV) file format. The BenchLink Meter software is used as a periodic trigger source for current measurements. To minimize spikes and ripple in our measurements software gathers five readings per trigger and averages the readings. In such setup the sample rate is approximately 33Hz. This sample frequency allows to identify software application level events and significant hardware events. The power levels are obtained from the current readings by multiplying current with battery voltage ($P = VI$). For measured scenarios the battery voltage is considered constant.

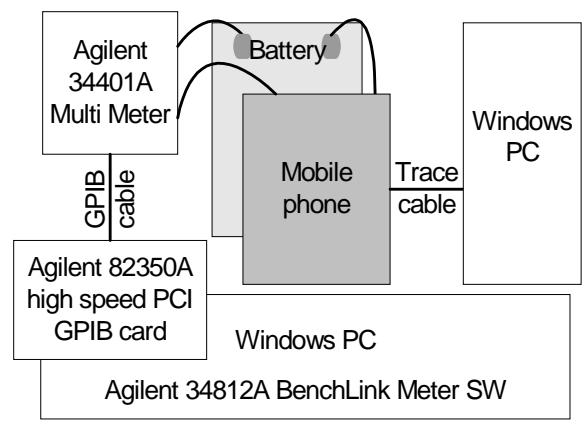

**Figure 4: Power measurement and tracing environment**

The software tracing is performed through a high-speed proprietary interface to the mobile phone. The timestamped trace is matched to the power measurements in Excel. The synchronization of traces and power measurements is performed using a synchronization event on the mobile device that is recorded in the software trace and produces a recognizable power spike in the power readings. The clock drift between two traces did not occur in our experiments. The experiments performed in this setup are described in the next section.

## 6 EXPERIMENTS

We have used the experimental setup to obtain traces and energy consumption data during the execution of different mobile phone applications including games, address book browser, calculator, SMS message edit and send, and phone calls. During the execution, the applications accessed different hardware components including a processor, a display system, an audio system, a cellular system and others. The power levels of these components were known and included in the model described in section 3. Some unmodeled hardware components were active as seen by the spikes in the graph.

The graph (Figure 5) shows the gathered measurement data and modeled power levels for a phone application as a function of time. As is visible from the graph, the model closely predicts the actual measurements of the power levels with the difference less than 5%. As seen from the graph, the energy consumption of software applications is determined by the hardware components accessed during the execution, so the energy consumption model fits the measured data.

The graph shows that different hardware devices have very different behaviors. While device 1 is on for most of the time, device 3 is turned on and off often and repeatedly. Devices also differ in how they are activated and deactivated. For example, parts of the keyboard hardware become active when keys are pressed, while the processor may be activated by periodic software activity or by hardware generated interrupts. On the other hand, some subsystems need to have power consuming "standby" modes that are enabled for the fast activation of their components.

The experiments above give the first evidence that software can be validated for the energy consumption. We are continuing to develop our model and obtaining additional measurements and traces to check the correctness of the phone applications.

## 7 RELATED WORK

There are numerous research projects investigating energy-efficient software implementation. Our work was inspired by the research at CMU on PowerScope [6]. As mentioned in section 4, PowerScope approach is one of the possible methods of gathering energy data. Researchers at CMU have found some errors in applications using PowerScope; however, there is no systematic work on using PowerScope for the software validation.

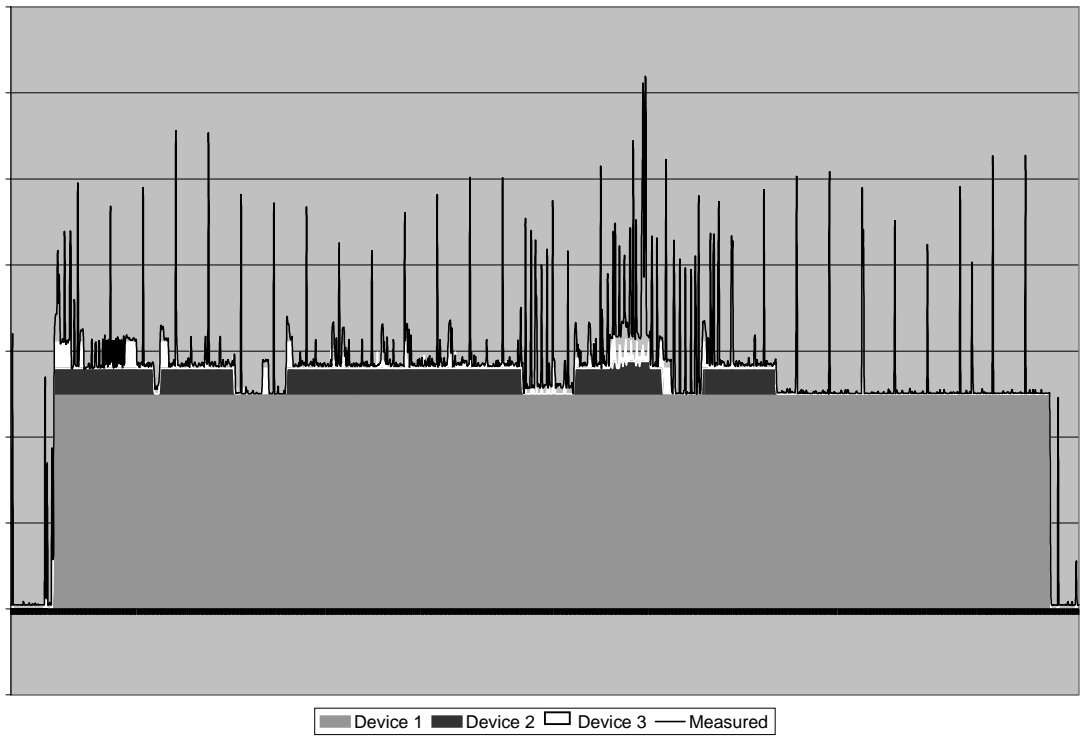

**Figure 5. Phone application power level graph**

Duke's Milly Watt project [5] focuses on providing a toolkit for energy measurements as well as developing an energy management API between hardware, operating system and applications. One of the tools will provide the program trace data together with power measurements. However, it is not currently available. Researchers from Duke University also proposed an energy consumption model for a Palm™ device [4], which they used only for the Palm™ device simulation and not for the validation. We formalized and extended their model using state machine diagrams and extended message sequence charts.

A research group at MIT working on power efficient systems has implemented JouleTrack [11] - a web based system for software energy profiling. JouleTrack currently simulates only the energy used by a processor in application execution. Such system could provide a detailed model and power function for a processor.

TTCN is defined by ISO/IEC [8] for the specification of tests for communication systems, and has been used for test case specification in various domains. The TTCN test case specification contains a stepwise sequence of message sends and replies including message parameters. The sequence can also include Boolean correctness conditions. The test outcome is specified at various levels of message sequences. The strength of TTCN is its capability to semi-formally specify communication protocol test sequences. Many protocol specifications can be directly used to define abstract test suites in TTCN. TTCN is widely adopted and integrated in specification, development, and testing tool suites such as Telelogic Tau. TTCN allows validating communication or other domain dependent protocols. However, the TTCN cannot be used for overall system validation using energy profile information

Real-time system modeling is a large research field. Our power consumption model has some similarities with timed transition systems [7] and modecharts [9]. Neither timed transition systems nor modecharts were previously used to model the energy consumption of the real-time systems.

## 8 CONCLUSIONS

This paper proposes the validation approach for the software quality requirement—its energy consumption. We have developed the power consumption model and test-case specifications that together with the software traces and power measurements allow validating energy requirements. By monitoring power levels and matching the measurements against modeled power levels we were able to validate the application power consumption. We continue developing a more detailed power consumption model to perform more exact software validation. We believe that the energy validation adds another dimension to the quality requirement validation.

## 9 ACKNOWLEDGEMENTS

We thank all the people from Nokia Mobile Phones who supported this research. We thank anonymous reviewers for valuable comments on this paper.

## 10 REFERENCES

[1] Agilent 34812A BenchLink Meter Software, Agilent Technologies, May 2001.

[2] Bluetooth, www.bluetooth.com, May 2001.

[3] G. Booch, J. Rumbaugh, I. Jacobson, *The Unified Modeling Language User Guide* (Addison-Wesley, 1999).

[4] T. Cignetti, K. Komarov, C. Ellis; Energy Estimation Tools for the Palm™, *Proceedings of ACM MSWiM 2000: Modeling, Analysis and Simulation of Wireless and Mobile Systems*, August 2000.

[5] C. Ellis; The Case for Higher-level Power Management, *Proceedings of the 7th Workshop on Hot Topics in Operating Systems (HotOS)*, Rio Rica, AZ, March 1999.

[6] J. Flinn, M. Satyanarayanan; PowerScope: A Tool for Profiling the Energy Usage of Mobile Applications, *Proceedings of the 2nd IEEE Workshop on Mobile Computing Systems and Applications*, New Orleans, Louisiana, February, 1999.

[7] T. A. Henzinger, Z. Manna, A. Pnueli. Temporal proof methodologies for timed transition systems. *Information and Computation*, 112, pp. 273-337, 1994.

[8] ISO/IEC 9646-3 (1991): "Information technology - Open Systems Interconnection – Conformance testing methodology and framework - Part 3: The Tree and Tabular Combined Notation (TTCN)".

[9] F. Jahanian, A. Mok, Modechart: A Specification Language for Real-Time Systems, *IEEE Transactions on Software Engineering*, vol. 20, no. 12, December 1994.

[10] D. F. Jerding, J. T. Stasko; The Information Mural: A technique for displaying and navigating large information spaces. *In Proceedings of the IEEE Visualization `95 Symposium on Information Visualization*, pages 43-50, Atlanta, GA, October 1995.

[11] A. Sinha, A. Chandrakasan; JouleTrack - A Web Based Tool for Software Energy Profiling, *Proceedings of the 38th Design Automation Conference*, Las Vegas, June 2001 (to appear), http://dry-martini.mit.edu/JouleTrack/